\documentclass[aps,prl,twocolumn,superscriptaddress,preprintnumbers,longbibliography,nofootinbib]{revtex4-2}

\usepackage{amssymb}
\usepackage{graphicx}
\usepackage{amsmath}
\usepackage{hyperref}
\usepackage{xspace}

\DeclareRobustCommand{\Fig}[1]{Fig.~\ref{#1}}
\DeclareRobustCommand{\Figs}[2]{Figs.~\ref{#1} and \ref{#2}}

\DeclareRobustCommand{\Eq}[1]{Eq.\,(\ref{#1})}

\DeclareRobustCommand{\Refs}[1]{Refs.\,\cite{#1}}

\usepackage{color}
\definecolor{darkblue}{rgb}{0,0,0.5}
\definecolor{darkred}{rgb}{0.5,0,0}
\definecolor{darkgreen}{rgb}{0,0.5,0}

\newcommand{\be}{\begin{equation}}
\newcommand{\ee}{\end{equation}}

\allowdisplaybreaks

\begin{document}

\preprint{}

\title{Understanding Jet Charge}

\author{Zhong-Bo Kang}
\email{zkang@ucla.edu}
\affiliation{Department of Physics and Astronomy, University of California, Los Angeles, CA 90095, USA}
\affiliation{Mani L. Bhaumik Institute for Theoretical Physics, University of California, Los Angeles, CA 90095, USA}
\affiliation{Center for Frontiers in Nuclear Science, Stony Brook University, Stony Brook, NY 11794, USA}
\author{Andrew J.~Larkoski}
\email{larkoski@ucla.edu}
\affiliation{Department of Physics and Astronomy, University of California, Los Angeles, CA 90095, USA}
\affiliation{Mani L. Bhaumik Institute for Theoretical Physics, University of California, Los Angeles, CA 90095, USA}
\author{Jinghong Yang}
\email{yangjh@umd.edu}
\affiliation{Department of Physics, University of Maryland, College Park, MD 20742, USA}
\affiliation{Department of Physics and Astronomy, University of California, Los Angeles, CA 90095, USA}

\begin{abstract} 
The jet charge is an old observable that has proven uniquely useful for discrimination of jets initiated by different flavors of light quarks, for example.  In this Letter, we propose an approach to understanding the jet charge by establishing simple, robust assumptions that hold to good approximation non-perturbatively, such as isospin conservation and large particle multiplicity in the jets, forgoing any attempt at a perturbative analysis.  From these assumptions, the jet charge distribution with fixed particle multiplicity takes the form of a Gaussian by the central limit theorem and whose mean and variance are related to fractional-power moments of single particle energy distributions.  These results make several concrete predictions for the scaling of the jet charge with the multiplicity, explaining many of the results already in the literature, and new results we validate in Monte Carlo simulation.  
\end{abstract}


\maketitle

As a conserved quantity unrelated to symmetries of spacetime, electric charge encodes information distinct from momentum of the mechanism of particle production in a high energy collision experiment, like the Large Hadron Collider (LHC).  On collimated streams of particles called jets, the total electric charge of all particles that compose a jet should, on average, be directly related to the electric charge of the short-distance quark or gluon that initiated the jet.  A definition of the jet charge $Q_\kappa$ robust to low-energy particles was proposed by Feynman and Field \cite{Field:1977fa} where
\begin{align}
Q_\kappa \equiv  \sum_{i\in J} z_i^\kappa\, Q_i\,,
\end{align}
where the sum runs over all particles $i$ in the jet $J$, $Q_i$ is the electric charge of particle $i$ in units of the fundamental charge $e$, and $\kappa > 0$ is a parameter that is responsible for the infrared safety of the jet charge.  $z_i$ is the energy fraction of particle $i$ in the jet, appropriately defined for the particular collider environment. This observable has been measured historically \cite{Fermilab-Serpukhov-Moscow-Michigan:1979zgc,Erickson:1979wa,Berge:1980dx,Aachen-Bonn-CERN-Munich-Oxford:1981lfk,Aachen-Bonn-CERN-Munich-Oxford:1982riw,EuropeanMuon:1984xji,Amsterdam-Bologna-Padua-Pisa-Saclay-Turin:1981hcw,TASSO:1990gda,DELPHI:1991mqi,ALEPH:1991fba,OPAL:1992jsm,OPAL:1994xvz,SLD:1994wsf,DELPHI:1996hzb,ALEPH:1997agc,CDF:1999jfn,L3:1999dig,OPAL:2000rnf,DELPHI:2001kqh,D0:2006kee,CDF:2013tpw,CDF:2013tpw}, as well as at both ATLAS and CMS experiments at the LHC \cite{ATLAS:2013mkl,CMS:2014rsx,ATLAS:2015rlw,CMS:2017yer,CMS:2020plq}.

Despite being extensively studied experimentally, the jet charge is not both infrared and collinear (IRC) safe and cannot be predicted within the perturbation theory of quantum chromodynamics (QCD) exclusively.  Many modern studies have analyzed the jet charge within the context of Monte Carlo event simulation or developed theoretical techniques for predictions of infrared but not collinear safe observables \cite{Krohn:2012fg,Waalewijn:2012sv,Chang:2013iba,Elder:2017bkd,Chen:2019gqo,Li:2019dre,Chen:2020vvp,Kang:2020fka,Kang:2021ryr,Li:2021zcf,Jaarsma:2022kdd,Chen:2022muj,Chen:2022pdu,Lee:2022kdn,Li:2021uww,Li:2023tcr}, but any such theoretical analysis requires significant input of non-perturbative data that cannot be predicted from first principles.  Ref.~\cite{Waalewijn:2012sv} pioneered the development of non-linear evolution equations that describe the perturbative scale dependence of the complete distribution of jet charge.  This formalism does enable identification of several predictions of the jet charge, especially related to its low moments and optimal choices of the parameter $\kappa$.

In this Letter, we construct a non-perturbative theoretical starting point for understanding the jet charge and to make concrete, robust predictions that make no reference to a short-distance description.  Other than studies within the context of simulated data, we are unaware of analyses of jet charge that forgo any attempt at an understanding based in perturbation theory.  To accomplish this, we present a set of assumptions that are guaranteed to be a good approximation non-perturbatively and from which calculations can be performed, similar to the approach of Ref.~\cite{Larkoski:2021hee}.  We believe that these assumptions are the simplest, minimal set and from them more details can be added, like including more flavors of quarks or incorporating short-distance correlations.  

The assumptions we use in this Letter are:
\begin{enumerate}
\item Particles (hadrons) in the jet are produced through identical, independent processes.
\item The multiplicity of particles in the jet $N$ is large.
\item The only particles are the pions: $\pi^+$, $\pi^-$, and $\pi^0$.
\item SU(2) isospin of the pions is an exact symmetry.
\end{enumerate}
These assumptions immediately enable us to write down the functional form of the probability distribution for the jet charge $Q_\kappa$ conditioned on the particle multiplicity $N$ of the jet.  The first two assumptions imply the central limit theorem and the jet charge distribution is Gaussian distributed.  The mean value $\langle Q_\kappa\rangle$ of the Gaussian is
\begin{align}
\langle Q_\kappa\rangle = N\langle z^\kappa Q\rangle\,,
\end{align}
where $\langle z^\kappa Q\rangle$ is the expectation value of the product of a single particle's energy fraction $z^\kappa$ and electric charge $Q$.  The variance $\sigma_\kappa^2$ is
\begin{align}
\sigma_\kappa^2\equiv \sum_{i\in J}z_i^{2\kappa}Q_i^2 = \sum_{\text{charged }i\in J}z_i^{2\kappa}\,,
\end{align}
assuming that $\langle Q_\kappa\rangle = 0$. From the third assumption, the squared electric charge of any particle in the jet is 0 or 1, and so only the charged particles contribute to the width.  Assuming exact isospin, the mean energy fraction $\langle z^{2\kappa}\rangle$ is identical for any of the pions and the number of charged pions is $2/3$ of the total multiplicity $N$.  The variance of the jet charge distribution can be expressed as
\begin{align}\label{eq:variance}
\sigma_\kappa^2 = \frac{2}{3}N\langle z^{2\kappa}\rangle\,,
\end{align}
where the expectation value of the energy fraction $\langle z^{2\kappa}\rangle$ is evaluated over all particles in the jet, charged or neutral.  By the translation-invariance of the variance, \Eq{eq:variance} holds for any value of the mean $\langle Q_\kappa\rangle$.

The probability distribution for the jet charge conditioned on the particle multiplicity in the jet is then
\begin{align}
p(Q_\kappa|N) = \frac{\exp\left[-\frac{\left(Q_\kappa - N\langle z^\kappa Q\rangle\right)^2}{\frac{4}{3}N\langle z^{2\kappa}\rangle}\right]}{\sqrt{2\pi \frac{2}{3}N\langle z^{2\kappa}\rangle}}\,.
\end{align}
This distribution depends on two moments, $\langle z^\kappa Q\rangle$ and $\langle z^{2\kappa}\rangle$, that can in principle be calculated from non-perturbative fragmentation functions \cite{Collins:1981uk,Collins:1981uw} or its scale dependence determined through a factorization theorem.  Again, we remain ignorant as to any short-distance description, and demonstrate the robust consequences of this result.

The first system to which we apply this analysis of jet charge is on an inclusive jet sample, in which the flavor (and hence the electric charge) of the parton that initiated the jet is random.  As such, and assuming exact isospin, the expectation value of the distribution of the jet charge vanishes, $\langle z^\kappa Q\rangle = 0$.

The expectation value $\langle z^{2\kappa}\rangle$, and therefore the variance $\sigma_\kappa^2$, is determined from the single-particle energy fraction distribution conditioned on the particle multiplicity in the jet, $p(z|N)$.  This moment is
\begin{align}
\langle z^{2\kappa}\rangle = \int_0^1 dz\, z^{2\kappa}\, p(z|N)\,.
\end{align}
Because the sum of momentum fractions of all $N$ particles in the jet must be 1, the expectation value of $z$ is
\begin{align}
\langle z\rangle = \int_0^1 dz\, z\, p(z|N) = \frac{1}{N}\,.
\end{align}
We can express the distribution $p(z|N)$ in a moment expansion about its mean, where
\begin{align}
p(z|N) = \delta\left(z-\frac{1}{N}\right) + \frac{\sigma_z^2}{2}\delta''\left(z-\frac{1}{N}\right) +\cdots\,.
\end{align}
Here, $\sigma_z^2$ is the variance of the distribution $p(z|N)$ and $\delta''(x)$ is the second derivative of the $\delta$-function.  While we cannot say much about the variance $\sigma_z^2$, it is non-negative and bounded from above, $0\leq \sigma_z^2< 1/N$, as $z\in[0,1]$ and $\langle z\rangle = 1/N$ \cite{doi:10.1080/00029890.2000.12005203}.  However, the variance is maximized by the pathological energy fraction distribution that consists of a sum $\delta$-functions at $z = 0,1$, with weights such that $\langle z\rangle=1/N$.  For any sufficiently smooth energy fraction distribution $p(z|N)$ the variance scales like $\sigma_z^2\sim 1/N^2$, which is what we assume in the following. We assume that higher orders in this moment expansion are small, and so are suppressed in the ellipses.

The variance of the jet charge distribution is then
\begin{align}\label{eq:jcvar}
\sigma_\kappa^2 
&= \frac{2}{3}N^{1-2\kappa}\left(1+\kappa(2\kappa-1)\sigma_z^2\, N^2+\cdots\right)\,.
\end{align}
The first two terms in this expansion should dominate for small $\kappa(2\kappa-1)\sigma_z^2\, N^2$, as $\sigma_z^2 N^2$ scales like $N^0$ by the smoothness assumption.  Note that this is guaranteed to be small if $\kappa$ is near $0.5$, which is the region in which measurements have been made.  Further, because $\sigma_z^2\geq 0$ and IR safety requires $\kappa > 0$, the sign of the second term in the moment expansion is exclusively determined by the value of $2\kappa-1$.  Finite width effects tend to decrease (increase) the jet charge width for $\kappa<0.5$ ($\kappa > 0.5$).

This expression for the jet charge width immediately makes the following predictions:
\begin{itemize}

\item For fixed particle multiplicity $N$, the jet charge distribution narrows as $\kappa$ increases.

\item For $\kappa < 0.5$ ($\kappa > 0.5$), the width of the jet charge distribution increases (decreases) as multiplicity $N$ increases, at a rate just slower than $N^{1-2\kappa}$.

\item For $\kappa = 0.5$, the width of the jet charge distribution is largely independent of the multiplicity $N$.
\end{itemize}
In all recent jet charge measurements \cite{ATLAS:2015rlw,CMS:2017yer,CMS:2020plq}, the narrowing of the jet charge distribution as $\kappa$ increases is directly observed.  Also, \Refs{Krohn:2012fg,Waalewijn:2012sv} demonstrated that the jet charge with $\kappa = 0.5$ has reduced dependence on the jet energy compared to other values of $\kappa$, which is consistent with the third prediction of this analysis.  However, the latter two predictions are challenging to discern in experimental data, because the jet charge distributions have as-of-yet not been binned in particle multiplicity.  

Given these concrete predictions for the behavior of the jet charge distribution, we test them in event simulation.  We generate $pp\to jj$ events at 5.02 TeV center-of-mass collisions in Pythia 8.303 \cite{Bierlich:2022pfr}.  Events are analyzed in Rivet 3.1.4 \cite{Bierlich:2019rhm} and its internal FastJet  \cite{Cacciari:2011ma} distribution. We find jets with a radius of $R = 0.4$ with the anti-$k_T$ algorithm \cite{Cacciari:2008gp}, and demand that the transverse momentum of the jets satisfies $p_\perp > 120$ GeV.  We only calculate the jet charge on the most central jet, with smallest absolute value of pseudorapidity.

\begin{figure}[t]
\includegraphics[width=0.45\textwidth]{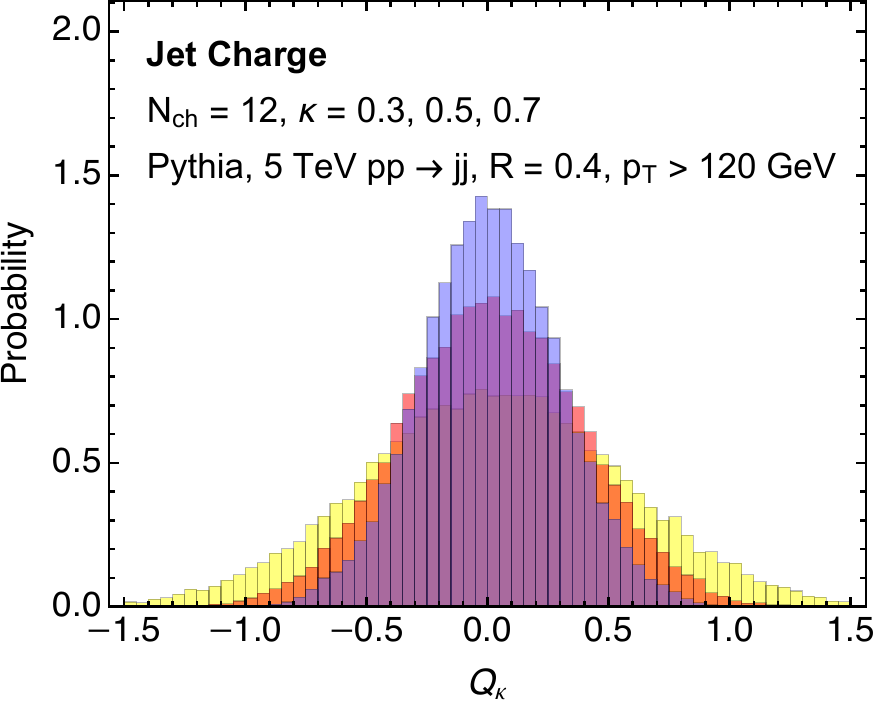}
\caption{\label{fig:jetcharge12}Histograms of the jet charge distribution with fixed charged particle multiplicity $N_\text{ch}=12$ for $\kappa=0.3,0.5,0.7$ in order of decreasing width.}
\end{figure}

\begin{figure}[t!]
\includegraphics[width=0.45\textwidth]{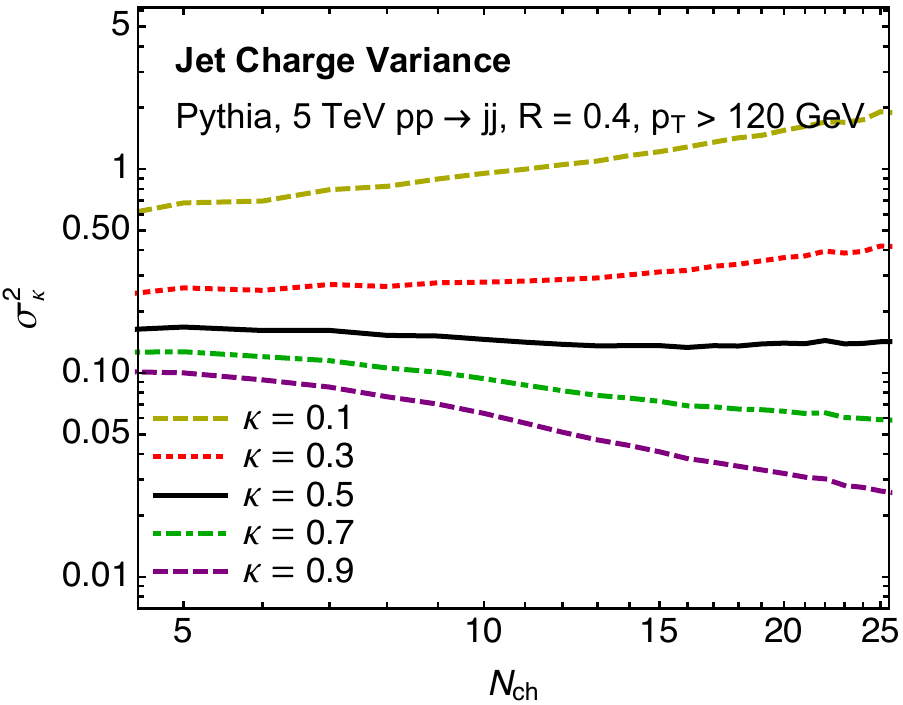}
\caption{\label{fig:varplot}Plot of the jet charge variance $\sigma_\kappa^2$ as a function of charged particle multiplicity, for $\kappa = 0.1,0.3,0.5,0.7,0.9$ from the top curve to bottom.}
\end{figure}

We validate our predictions on this simulated data in \Figs{fig:jetcharge12}{fig:varplot}.  In \Fig{fig:jetcharge12}, we plot the jet charge distribution on jets with charged particle multiplicity fixed to $N_\text{ch}=12$, which is the mode of the multiplicity distribution on this sample of jets.  As $\kappa$ increases, the distributions clearly become more narrow, as our scaling analysis predicts.  In \Fig{fig:varplot}, we plot the jet charge variance $\sigma_\kappa^2$ for several values of $\kappa$, as a function of $N_\text{ch}$.  The width for $\kappa = 0.5$ is constant to very good approximation, and the logarithm of the width increases (decreases) for smaller (larger) values of $\kappa$ approximately linearly in $\log N_\text{ch}$, exactly as predicted.  Further plots that validate our predictions are presented in Supplemental Material.

Jet charge is a promising observable for discrimination of jets initiated by different flavors of light quarks, with perhaps the most interesting scenario of separation of jets initiated by up versus down quarks \cite{Krohn:2012fg,Kang:2020fka}.  We assume here that a jet's partonic flavor is well-defined, with several jet flavor definitions developed recently \cite{Banfi:2006hf,Caletti:2022hnc,Caletti:2022glq,Czakon:2022wam,Gauld:2022lem}.   A recent machine learning study on up versus down quark jet discrimination \cite{Lee:2022kdn} demonstrated that there was essentially no useful kinematical information, and that exclusively the jet charge provided the separation power.  Within the context of our assumptions, we can understand many of the results of these and other analyses on this problem.

Unlike jet charge on inclusive jets, the expectation value of the jet charge distribution on jets initiated by up or down quarks is non-zero.  To evaluate the expectation value $\langle Q_\kappa\rangle$ on up or down quark jets, we assume that the mechanisms for production of particle electric charge and momentum are independent, and so
\begin{align}
\langle Q_\kappa\rangle = N\langle z^\kappa Q\rangle = N\langle z^\kappa\rangle\langle Q\rangle\,.
\end{align}
Of course, electric charge and momentum production are correlated perturbatively, so we expect this assumption to be approximately  accurate to the $\alpha_s\sim 10\%$ level. The average electric charge of a single particle in the jet $\langle Q\rangle$ is determined by the initiating particle because isospin conservation forbids the production of additional net charge.  For up and down jets, respectively, the mean single-particle charges are
\begin{align}
&\langle Q\rangle_u =\frac{2}{3}\, N^{-1}\,,
&\langle Q\rangle_d =-\frac{1}{3}\, N^{-1}\,.
\end{align}
The expectation values of the jet charge on up or down jets are
\begin{align}
&\langle Q_\kappa\rangle_u =\frac{2}{3}\langle z^\kappa\rangle\,, 
&\langle Q_\kappa\rangle_d =-\frac{1}{3}\langle z^\kappa\rangle\,.
\end{align}
Generalizing from SU(2) isospin to SU(3) flavor changes these net charges slightly, as well as introducing contributions from strange quarks \cite{Field:1977fa}, but we do not consider this here.

We assume that the expectation value $\langle z^\kappa\rangle$ is identical for all quark jets and can be calculated with the moment expansion of the distribution $p(z|N)$ used in the previous section.  We find
\begin{align}
\langle z^\kappa\rangle 
&= N^{-\kappa}\left(
1+\frac{\kappa}{2}(\kappa-1)\sigma_z^2N^2+\cdots\,.
\right)\,.
\end{align}  
For all $0<\kappa < 1$, finite width effects tend to decrease the magnitude of the average jet charge.  For the width of the jet charge distribution on these up and down quark jets, we again assume that their width is identical and can be expressed in the same form as in Eq.~\eqref{eq:jcvar}.  The width $\sigma_z^2$ on up and down jets may be different than for the inclusive jet sample because of the presence of gluon-initiated jets.

To maximize discrimination power between up and down jets with jet charge, we choose a value of $\kappa$ such that the width $\sigma_\kappa^2$ of the distributions is minimized compared to the difference squared of the expectation values of the up and down jet charges.  Using the notation of Ref.~\cite{Waalewijn:2012sv}, we consider the quantity
\begin{align}
\eta(\kappa) \equiv \frac{\left(\langle Q_\kappa\rangle_u-\langle Q_\kappa\rangle_d\right)^2}{\sigma_\kappa^2} =\frac{\langle z^\kappa\rangle^2}{\frac{2}{3}N\langle z^{2\kappa}\rangle}\,,
\end{align}
and determine the optimal value $\kappa_*$ as where $\eta(\kappa)$ is maximized.  This ratio of the moments is
\begin{align}
\eta(\kappa)=\frac{\langle z^\kappa\rangle^2}{\frac{2}{3}N\langle z^{2\kappa}\rangle} =\frac{1}{\frac{2}{3}N}\left( 1-\kappa^2\sigma_z^2 N^2+\cdots\right)\,.
\end{align}
This predicts that the optimal value of $\kappa$ that should be chosen for maximizing discrimination power between up and down quark jets is $\kappa_* \to 0$. In this limit, the jet charge loses its IR safety and is very sensitive to contributions from arbitrarily soft particles, so taking the strict $\kappa_* = 0$ limit is not optimal.  This is a similar feature to the analytic observation that recoil-free, IRC safe observables with the weakest angular weighting possible provide optimal discrimination between jets initiated by quarks and gluons \cite{Larkoski:2013eya}.  As in that case, to ensure IRC safety, the angular weighting cannot strictly vanish, and here also the energy weighting cannot disappear.  This does predict that the discrimination power of the jet charge is improved with small values of $\kappa$ (down to some minimum imposed by IR safety), which has been observed in several previous simulation studies, e.g., \Refs{Krohn:2012fg,Waalewijn:2012sv,Lee:2022kdn}.

The form of the mean-width ratio $\eta(\kappa)$ also informs the behavior of the jet charge's discrimination power as a function of jet multiplicity $N$.  As the multiplicity increases, $\eta(\kappa)$ decreases, corresponding to degraded discrimination power.  Further, particle multiplicity increases as the energy of the jet increases, so we expect that the power of jet charge to identify jets initiated by up or down quark jets also degrades at higher energies.  In \Refs{Krohn:2012fg,Waalewijn:2012sv}, the energy dependence of the mean and width of the jet charge distribution was calculated from a perturbative factorization theorem.  From these results, it was observed that both the mean and width decrease as jet energy increases, but that the mean decreased at a faster rate than the width.  For up and down quark jet discrimination, this implies that $\eta(\kappa)$ decreases as the jet energy increases, consistent with our scaling analysis.

We can now construct the joint probability distribution of the jet charge and particle multiplicity.  Assuming that the normalized multiplicity distributions of up and down quark jets are identical, $p_u(N) = p_d(N)\equiv p(N)$, the joint probability distributions are
\begin{align}
&p_u(Q_\kappa,N) = p_u(Q_\kappa|N)\,p(N)\,,\\
&p_d(Q_\kappa,N) = p_d(Q_\kappa|N)\,p(N)\,.
\end{align}
By being differential in both jet charge and multiplicity, we can potentially construct a discrimination observable that is more powerful than jet charge alone.  By the Neyman-Pearson lemma \cite{Neyman:1933wgr}, the optimal observable ${\cal O}$ for binary discrimination is (monotonic in) the logarithm of the likelihood ratio, where 
\begin{align}
{\cal O}\equiv \log{\cal L} = \log\frac{p_u(Q_\kappa,N)}{p_d(Q_\kappa,N)} = \log\frac{p_u(Q_\kappa|N)}{p_d(Q_\kappa|N)}\,.
\end{align}
The explicit expression for this observable in terms of $Q_\kappa$ and $N$ can be established using the Gaussian form of the distributions.  Working to lowest order in the moment expansion, this observable is
\begin{align}\label{eq:loglikeanal}
{\cal O} = \frac{3}{2}
N^{-1+\kappa} Q_\kappa - \frac{N^{-1}}{4}\,.
\end{align}
This is not monotonically related to the jet charge $Q_\kappa$, but instead contains non-trivial correlations between multiplicity and jet charge.  We expect that this observable is a better discriminant between jets initiated by up and down quarks than the jet charge alone.  More generally, the form of this observable demonstrates that discrimination power can be improved by performing measurements simultaneously differential in both jet charge and particle multiplicity.

To summarize, the predictions for up and down quark jet discrimination are:
\begin{itemize}
\item As $\kappa\to 0$, the discrimination power of the jet charge improves down to some minimal value below which infrared effects become uncontrolled.

\item The discrimination power of the jet charge decreases as the particle multiplicity increases.  

\item There is useful discrimination information in the joint distribution of jet charge and particle multiplicity.
\end{itemize}

\begin{figure}[t!]
\includegraphics[width=0.45\textwidth]{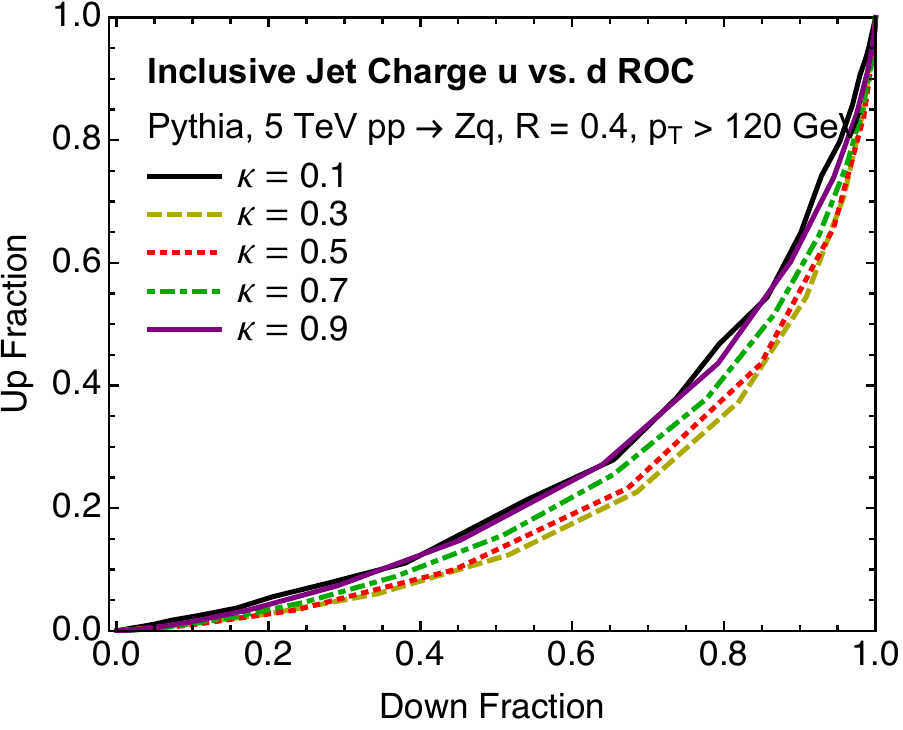}
\caption{\label{fig:roc_inc}Plot of the ROC curve for up versus down quark discrimination with the jet charge observable, for values of exponent $\kappa = 0.1,0.3,0.5,0.7,0.9$.}
\end{figure}

We generate $pp\to Z+u$ and $pp\to Z+d$ at leading order in MadGraph 3.4.1 \cite{Alwall:2014hca} and have the $Z$ boson decay exclusively to neutrinos, which are removed before jet finding.  These events are passed to Pythia for the same parton shower and jet finding analysis as for the inclusive QCD jet samples.  In \Fig{fig:roc_inc}, we plot the receiver operating characteristic (ROC) curve representing the discrimination power of the jet charge, inclusive in particle multiplicity, for different values of the exponent $\kappa$.  As predicted from the scaling analysis, there is better discrimination power at smaller $\kappa$; that is, a higher fraction of down quark jets can be isolated for the same fraction of contamination of up jets.  However, at very small $\kappa \sim 0.1$, IR effects dominate, and the discrimination power is degraded.

\begin{figure}[t!]
\includegraphics[width=0.45\textwidth]{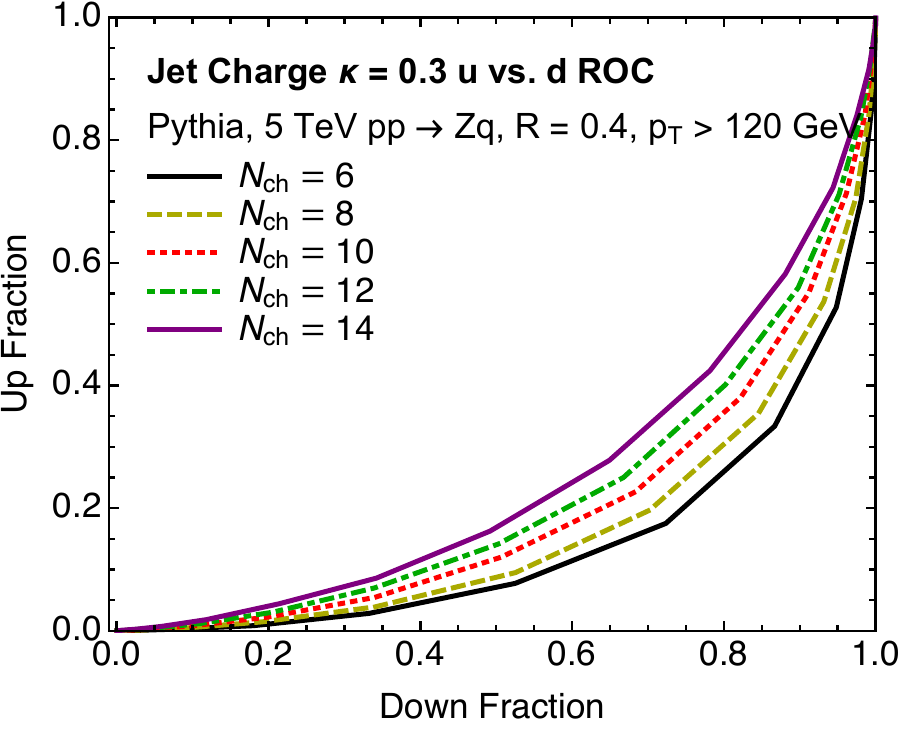}
\caption{\label{fig:roc_nch}Plot of ROC curves for up versus down quark discrimination with the jet charge with $\kappa = 0.3$, separated into different values of charged particle multiplicity, $N_\text{ch}=6,8,10,12,14$.}
\end{figure}

\begin{figure}[t!]
\includegraphics[width=0.45\textwidth]{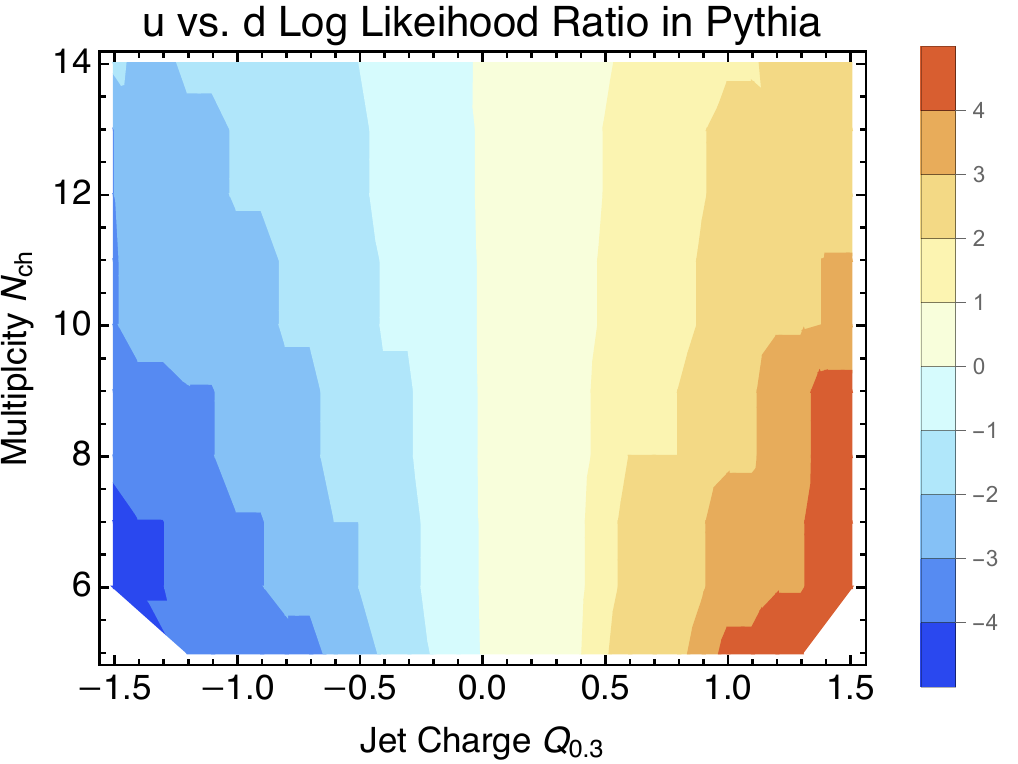}
\caption{\label{fig:loglike}Contour plot of the logarithm of the likelihood ratio for up versus down quark discrimination, differential in the jet charge with $\kappa = 0.3$ and the charged particle multiplicity.  Positive (negative) values represent regions dominated by up (down) quark jets.}
\end{figure}

The relationship between the jet charge and charged particle multiplicity is explored in \Figs{fig:roc_nch}{fig:loglike}.  In \Fig{fig:roc_nch}, we plot the ROC curve for the jet charge with $\kappa=0.3$ split into different values of particle multiplicity, and as multiplicity increases, the discrimination power decreases, as predicted.  In \Fig{fig:loglike}, we plot the logarithm of the likelihood ratio of the jet charge for $\kappa = 0.3$ versus the charged particle multiplicity.  If these observables were completely uncorrelated for discrimination, the contours would be perfectly vertical, but away from $Q_\kappa = 0$, the contours mix multiplicity and jet charge, demonstrating that discrimination can be improved by measuring both jet charge and multiplicity.  In the Supplemental Material, we show that contours of the analytic prediction for the logarithm of the likelihood ratio, Eq.~\ref{eq:loglikeanal}, agree well with the results from simulation in \Fig{fig:loglike}.

We have introduced simple, non-perturbative assumptions that form a foundation for understanding the jet charge observable.  These assumptions make concrete predictions especially with regards to the relationship between the jet charge and the particle multiplicity that are born out in simulation.  The central conclusion from this analysis is that future measurements of the jet charge should be binned in charged particle multiplicity, as this both makes scaling properties of the jet charge manifest as well as improve performance for discrimination problems.  We look forward to applications of this observable at the LHC and at future colliders like the Electron-Ion Collider.

Z.K. and A.L. are supported by the National Science Foundation under grant No.~PHY-1945471. J.Y.~thanks the UCLA College Honors program for the Honors Summer Research Stipend.  This work was supported in part by the UC Southern California Hub, with funding from the UC National Laboratories division of the University of California Office of the President.

\bibliography{jet_charge}

\end{document}


\title{Understanding Jet Charge: Supplemental Material}

\author{Zhong-Bo Kang}
\email{zkang@ucla.edu}
\affiliation{Department of Physics and Astronomy, University of California, Los Angeles, CA 90095, USA}
\affiliation{Mani L. Bhaumik Institute for Theoretical Physics, University of California, Los Angeles, CA 90095, USA}
\affiliation{Center for Frontiers in Nuclear Science, Stony Brook University, Stony Brook, NY 11794, USA}
\author{Andrew J.~Larkoski}
\email{larkoski@ucla.edu}
\affiliation{Department of Physics and Astronomy, University of California, Los Angeles, CA 90095, USA}
\affiliation{Mani L. Bhaumik Institute for Theoretical Physics, University of California, Los Angeles, CA 90095, USA}
\author{Jinghong Yang}
\email{yangjh@umd.edu}
\affiliation{Department of Physics, University of Maryland, College Park, MD 20742, USA}
\affiliation{Department of Physics and Astronomy, University of California, Los Angeles, CA 90095, USA}

\maketitle

\begin{figure}
\includegraphics[width=0.45\textwidth]{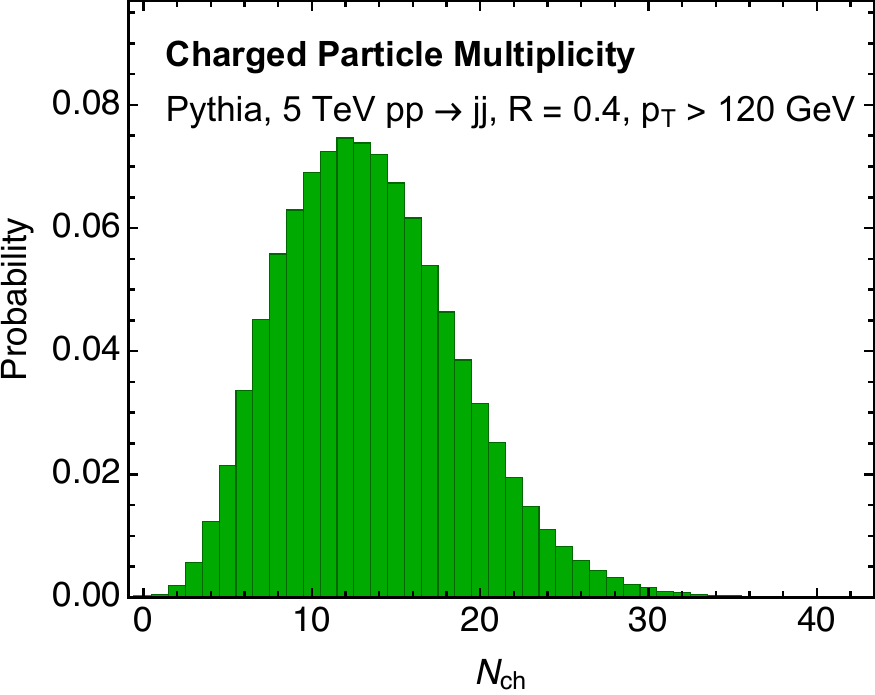}
 \ \ \includegraphics[width=0.45\textwidth]{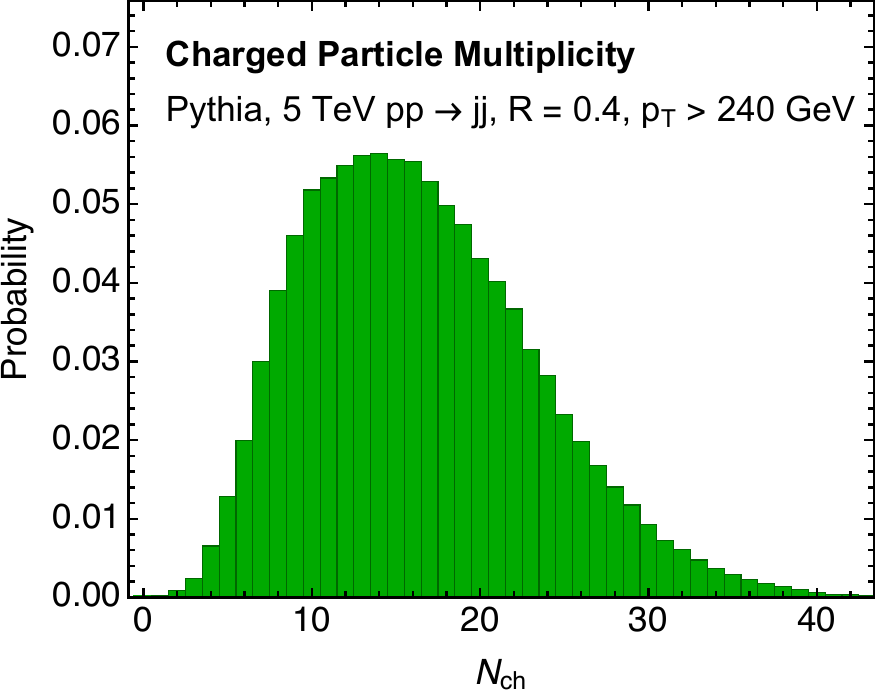}
\caption{\label{fig:mult} Plots of the charged particle multiplicity distributions for jets with $p_T > 120$ GeV (left) and $p_T > 240$ GeV (right).}
\end{figure}

Here, we present Supplemental Material for the article that further validates the predictions made from our scaling analysis.  We first plot the charged particle multiplicity distributions on the $pp\to jj$ samples in \Fig{fig:mult}, for jets with $p_T > 120$ GeV (left) and $p_T > 240$ GeV (right).  The mode of the distribution for jets with $p_T > 120$ GeV is $N_\text{ch}=12$, which is the value of multiplicity used in the main article.  Note that the charged particle multiplicity increases as the jet $p_T$ increases, which will have an effect on the jet charge distribution.

\begin{figure}
\includegraphics[width=0.45\textwidth]{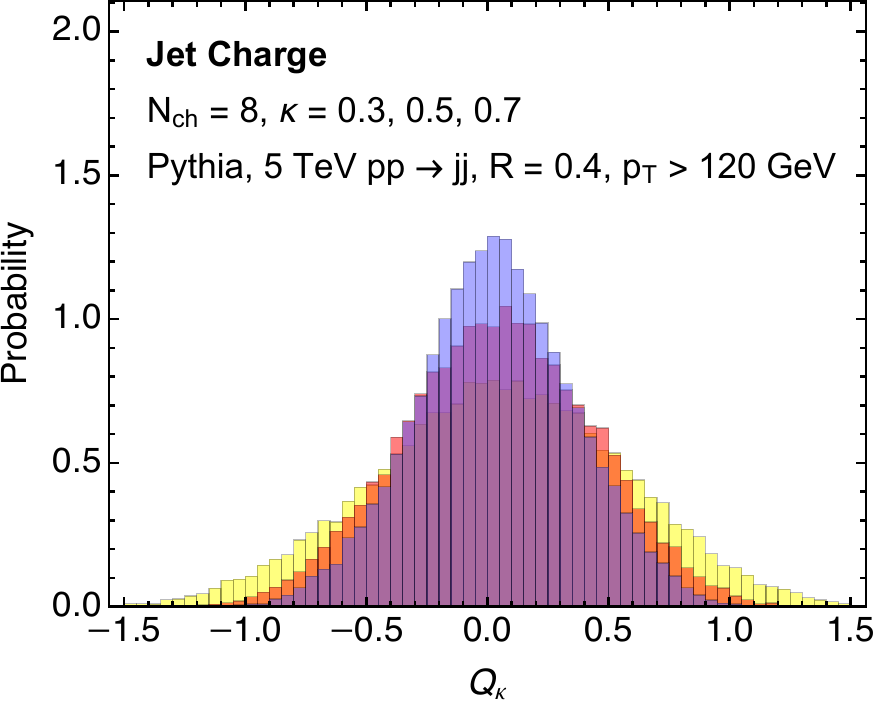}\ \includegraphics[width=0.45\textwidth]{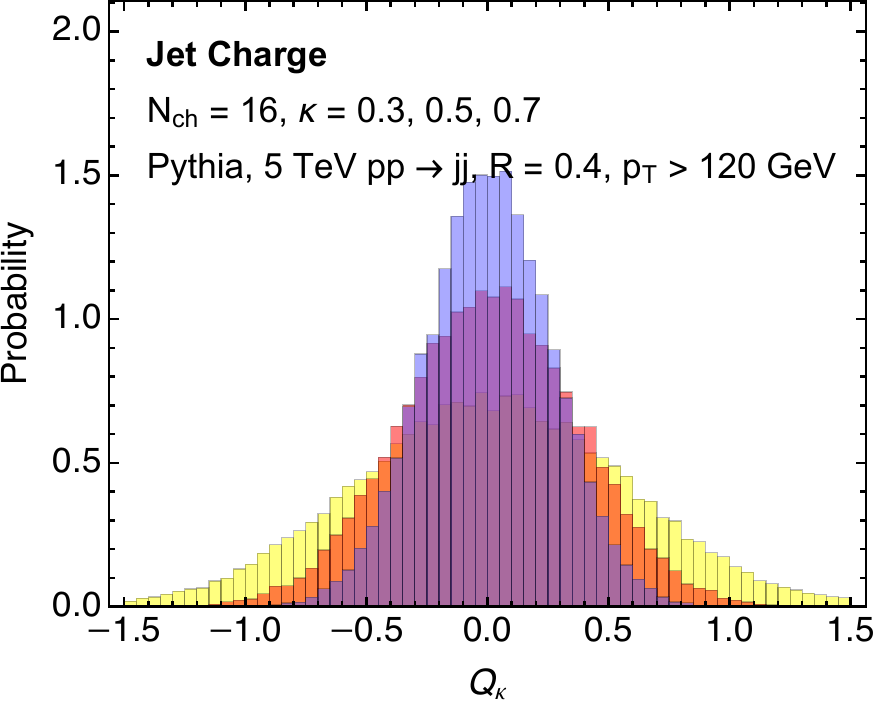}
\caption{\label{fig:jetcharge816}Plots of the jet charge distribution on the jet sample with $p_T > 120$ GeV with $\kappa =0.3,0.5,0.7$ for two different values of charged particle multiplicity: $N_\text{ch}=8$ (left) and $N_\text{ch}=16$ (right).}
\end{figure}

In \Fig{fig:jetcharge816}, we plot the jet charge distribution on the sample of jets with $p_T > 120$ GeV, with $\kappa = 0.3,0.5,0.7$ and for two different values of charged particle multiplicity, $N_\text{ch}=8, 16$.  This further illustrates the dependence of the width of the jet charge distribution on the multiplicity, as given by Eq.~(9) in the main text.

\begin{figure}
\includegraphics[width=0.45\textwidth]{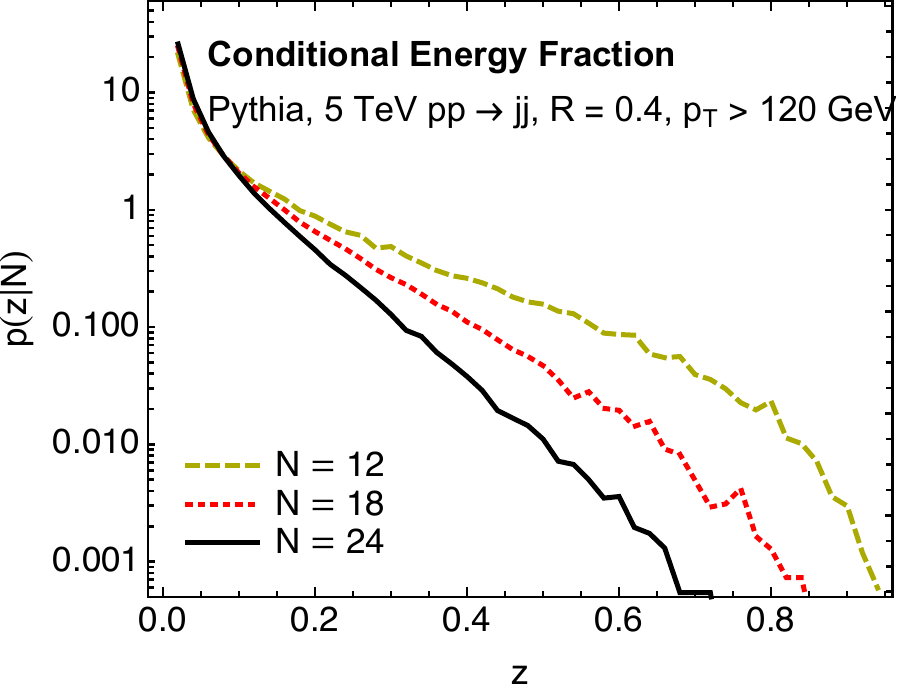}
\caption{\label{fig:fragfunc}Plot of the single-particle energy fraction distribution conditioned on total multiplicity $N$, $p(z|N)$, for $N = 12,18,24$.}
\end{figure}

In \Fig{fig:fragfunc}, we plot the single-particle energy fraction distribution conditioned on the total just multiplicity, $p(z|N)$, for three values of total multiplicity $N = 12,18,24$.  As expected, the distributions narrow as multiplicity increases.  Additionally, on these distributions we can validate our assumption that the variance $\sigma_z^2$ scales like $N^{-2}$.  This assumption was necessary to ensure that the moment expansion would converge and the first two terms be representative of the total distribution.  From these three distributions, we calculate the product of the variance and the squared multiplicity, $\sigma_z^2 N^2$.  We find
\begin{align}
&N = 12:\ \ \sigma_z^2 N^2 = 2.25\,,\\
&N = 18:\ \ \sigma_z^2 N^2 = 2.30\,,\\
&N = 24:\ \ \sigma_z^2 N^2 = 2.09\,.
\end{align}
These results have very little residual depedence on multiplicity, validating our assumption made in the main manuscript.  Additionally, we note that this scaling of the variance differs by about a factor of 2 from the result of $\sigma_z^2 N^2$ if $p(z|N)$ were an exponential distribution with mean $\langle z\rangle = 1/N$.

\begin{figure}
\includegraphics[width=0.45\textwidth]{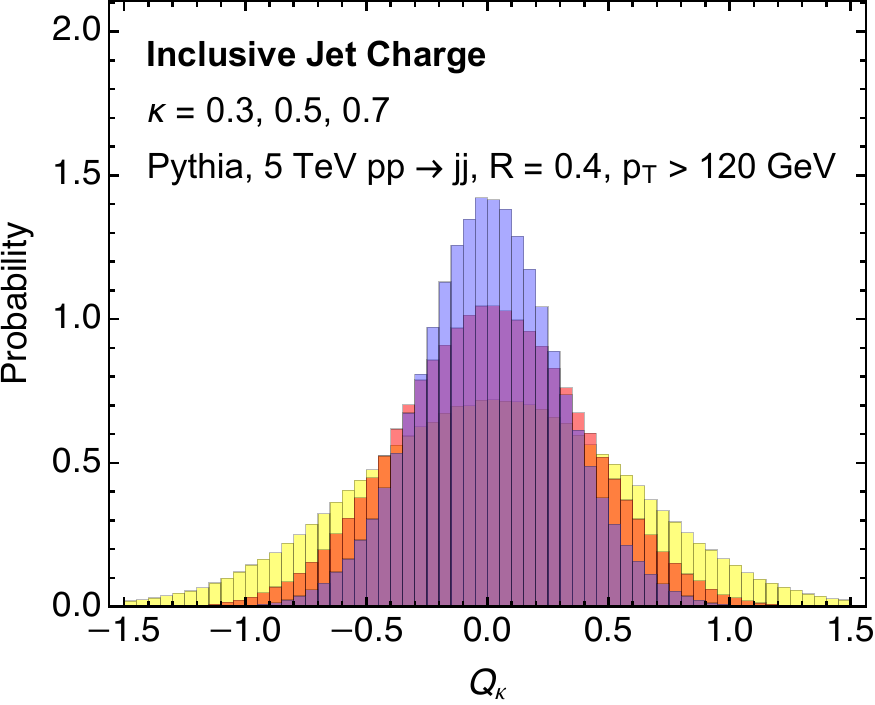}
\ \ \includegraphics[width=0.45\textwidth]{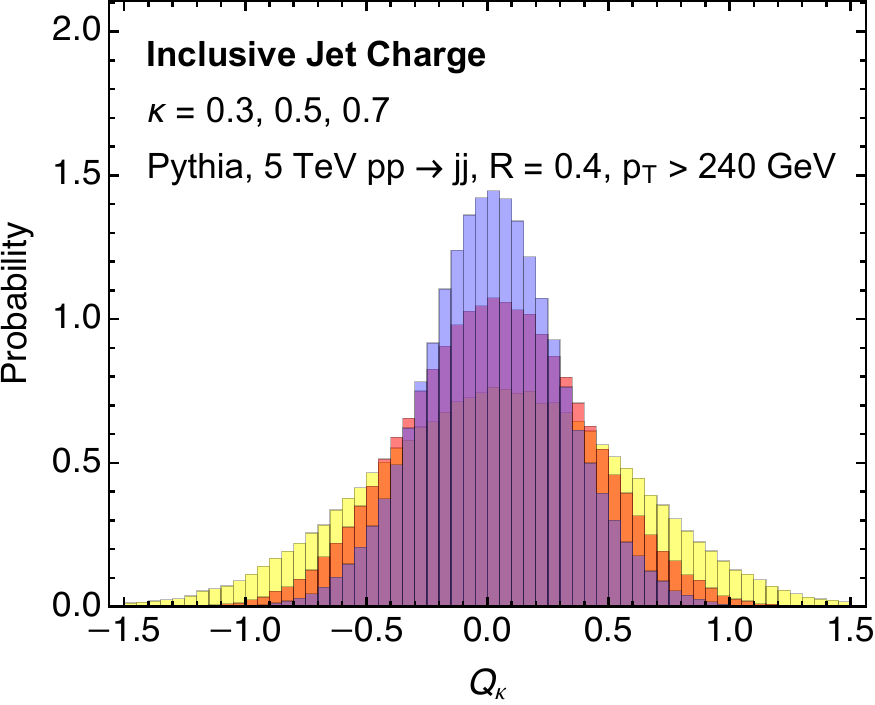}
\caption{\label{fig:incharge}Plots of the jet charge distributions inclusive over particle multiplicity for jets with $p_T > 120$ GeV (left) and $p_T > 240$ GeV (right).}
\end{figure}

\begin{figure}
\includegraphics[width=0.45\textwidth]{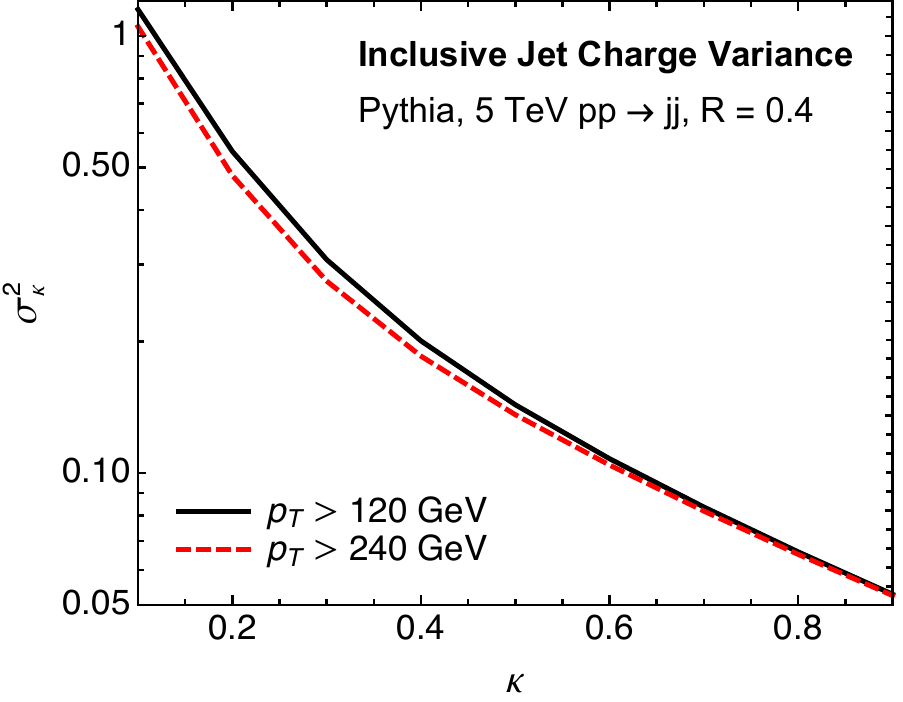} \ \ \includegraphics[width=0.45\textwidth]{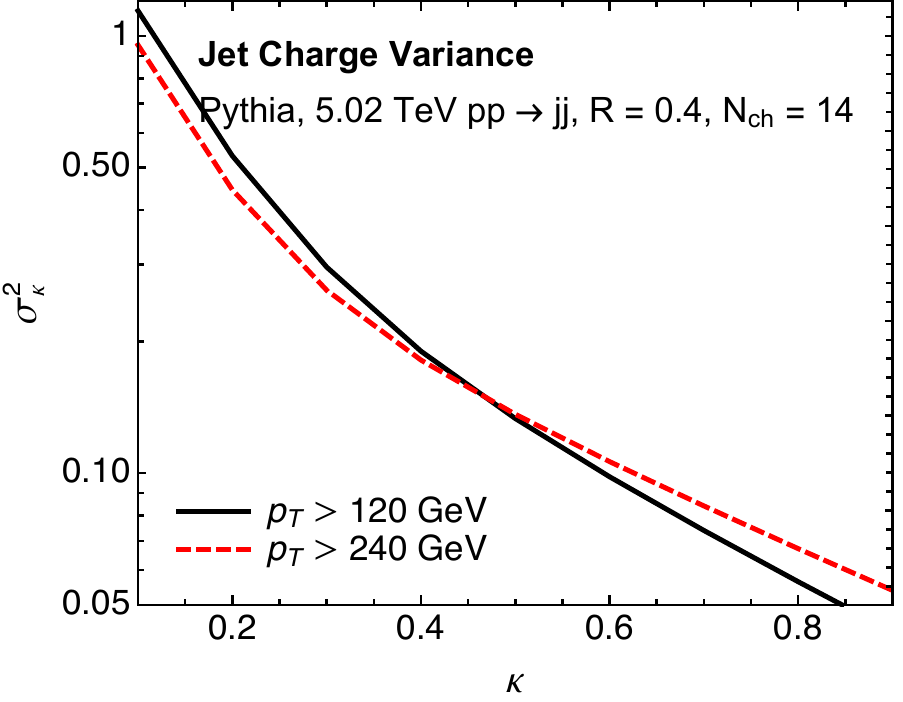}
\caption{\label{fig:var_comp}Plot comparing the variance of the inclusive (left) or $N_\text{ch} = 14$ (right) jet charge distribution as a function of exponent $\kappa$ for jets with $p_T > 120$ GeV (solid black) and $p_T > 240$ GeV (dashed red).}
\end{figure}

Because the multiplicity increases as jet $p_T$ increases, we expect that this framework predicts the dependence of the width on the $p_T$, as well.  First, in \Fig{fig:incharge}, we plot the inclusive jet charge for $\kappa = 0.3,0.5,0.7$ and $p_T > 120$ GeV and $p_T > 240$ GeV.  From these plots, there is very little difference between the distributions discernable by eye.  At left in \Fig{fig:var_comp}, we plot the variance of the jet charge as a function of $\kappa$ at the different jet $p_T$ cuts.  Here, we see that the higher $p_T$ sample is very slightly narrower than the lower $p_T$ sample over the entire range of $\kappa$.  This differs from our scaling analysis predictions, for which increasing the multiplicity increases (decreases) the width for $\kappa < 0.5$ ($\kappa > 0.5$).

To understand this result a bit more, at right in \Fig{fig:var_comp}, we plot the variance in these two jet $p_T$ samples as a function of $\kappa$, but with fixed charged particle multiplicity $N_\text{ch} = 14$.  Our scaling analysis would predict that these widths should be identical with the same multiplicity, but we see that they are not.  The mechanism of particle production in jets is $p_T$ dependent, and this fact is not accounted for in our simple analysis.  That is, the single particle energy fraction distribution $p(z|N)$ from which we calculated the variance of the jet charge depends on jet $p_T$.  This scale dependence should be calculable within perturbation theory, but is beyond our analysis here.  Nevertheless, these effects are relatively small compared to our scaling predictions, validating that they are a useful starting point for understanding the jet charge.

\begin{figure}
\includegraphics[width=0.45\textwidth]{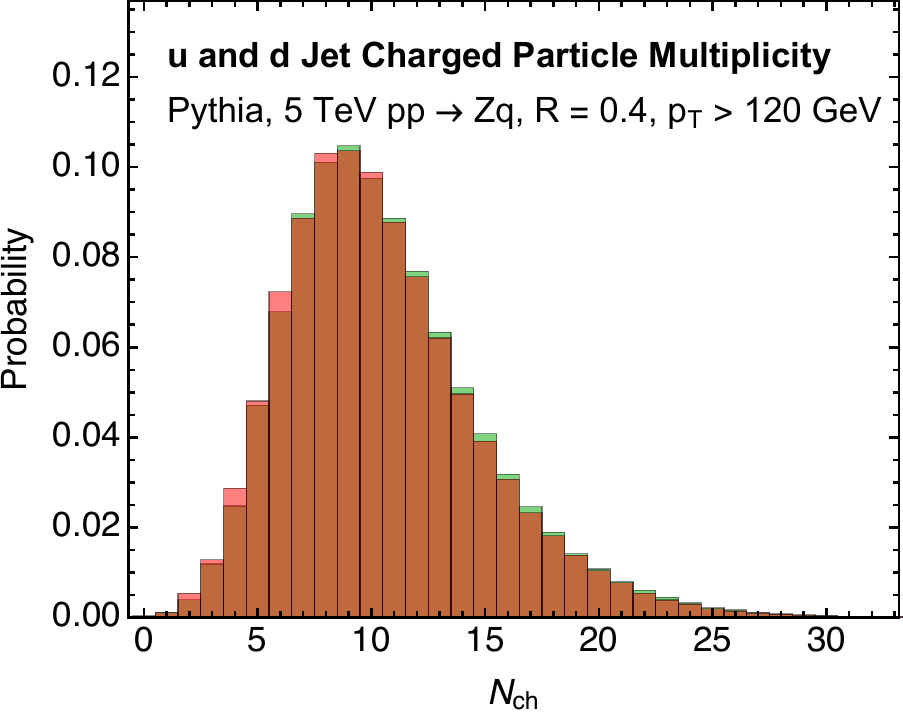}
 \ \ \includegraphics[width=0.45\textwidth]{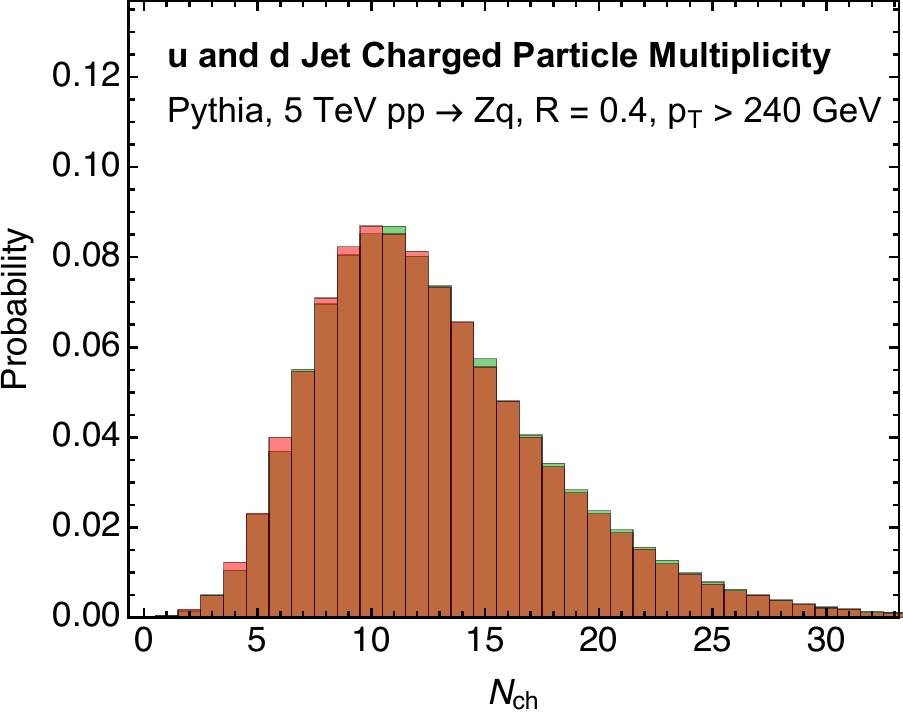}
\caption{\label{fig:mult_uvsd} Plots of the charged particle multiplicity distributions for up (green) and down (red) jets with $p_T > 120$ GeV (left) and $p_T > 240$ GeV (right).}
\end{figure}

Moving on to the up versus down jet analysis and predictions, we plot the charged particle multiplicity in jets with both $p_T > 120$ GeV and $p_T > 240$ GeV in \Fig{fig:mult_uvsd}.  As assumed in the main body of the Letter, the charged particle multiplcity distributions of up and down jets are nearly identical for a given $p_T$ selection.  Also, as expected, the particle multiplicity in the jets increases as the $p_T$ cut increases.

\begin{figure}
\begin{center}
\includegraphics[width=0.45\textwidth]{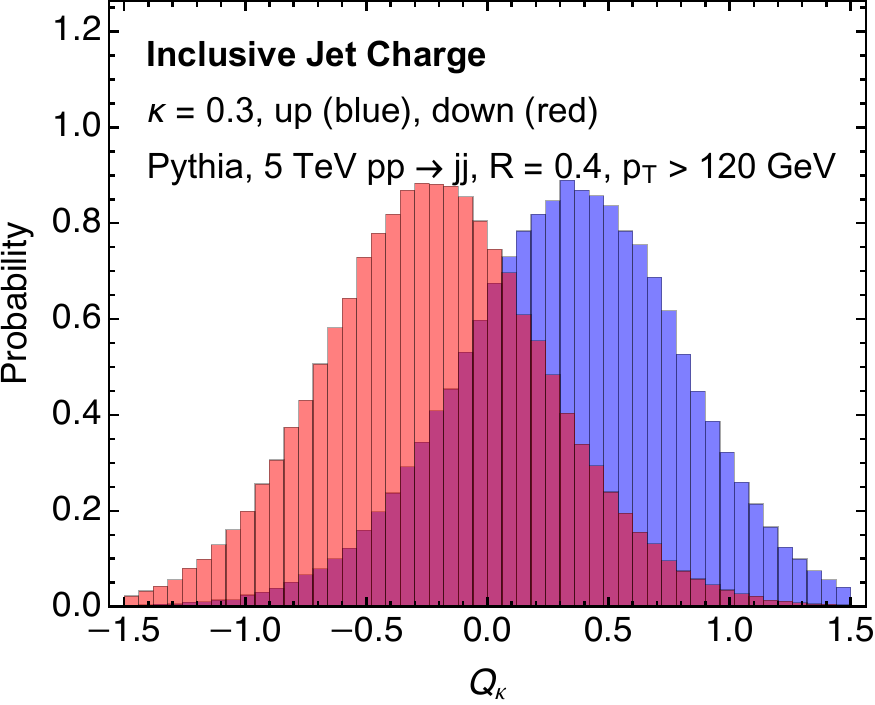}\ \includegraphics[width=0.45\textwidth]{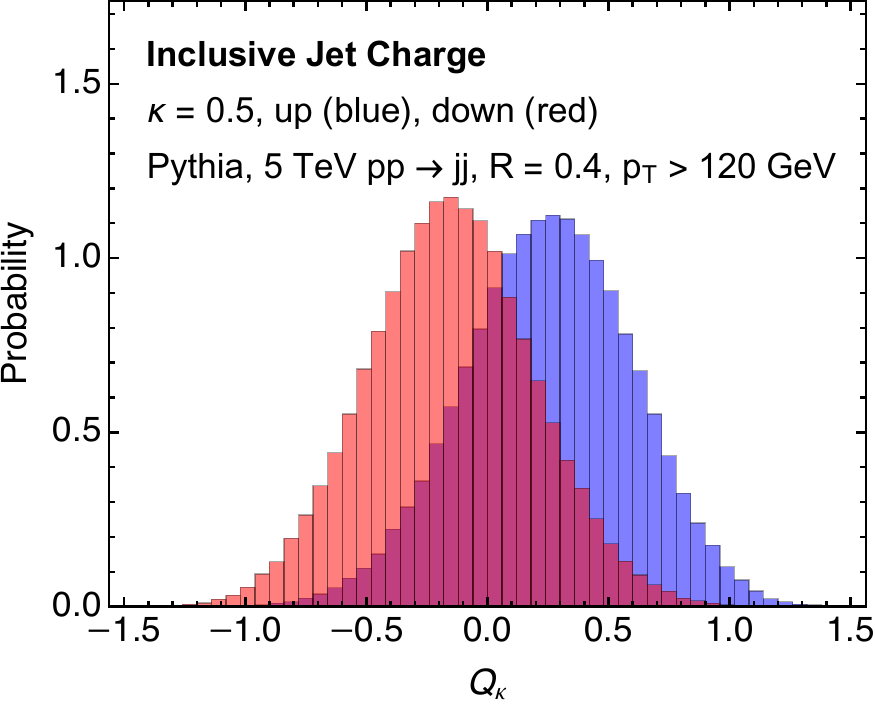}\\
 \includegraphics[width=0.45\textwidth]{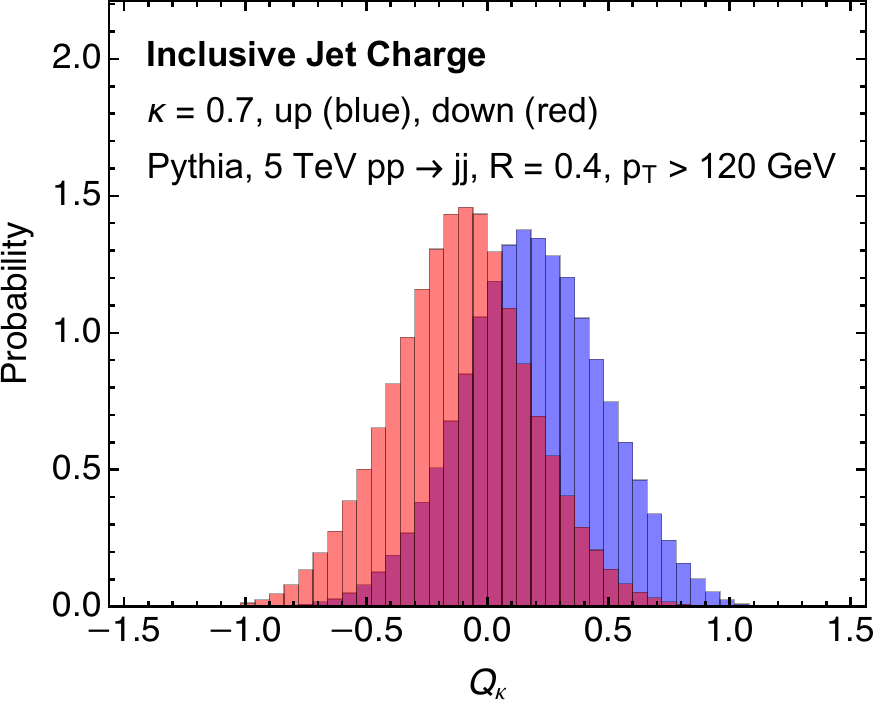}
\caption{\label{fig:jetcharge_uvsd}Plots of the jet charge distribution on up and down quark jet samples with $p_T > 120$ GeV with $\kappa =0.3,0.5,0.7$, inclusive over charged particle multiplicity.}
\end{center}
\end{figure}

In \Fig{fig:jetcharge_uvsd}, we plot the jet charge distributions on the up and down jets, inclusive over multiplicity and for $\kappa = 0.3,0.5,0.7$.  As our scaling analysis predicts, both the mean and variance of the distributions decreases as $\kappa$ increases.  

\begin{figure}
\begin{center}
\includegraphics[width=0.45\textwidth]{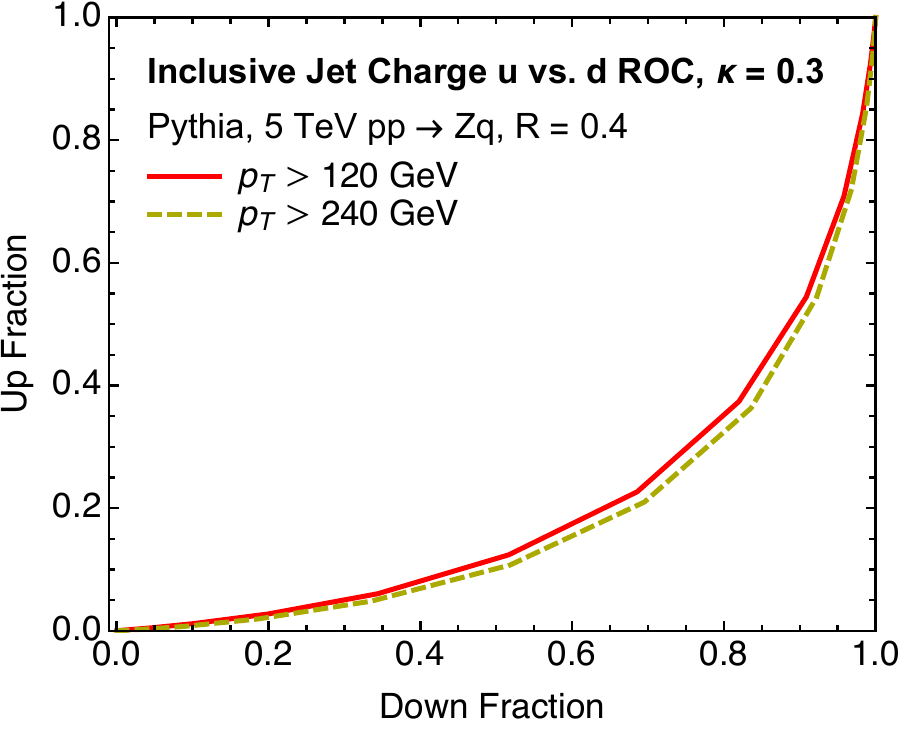}\ \includegraphics[width=0.45\textwidth]{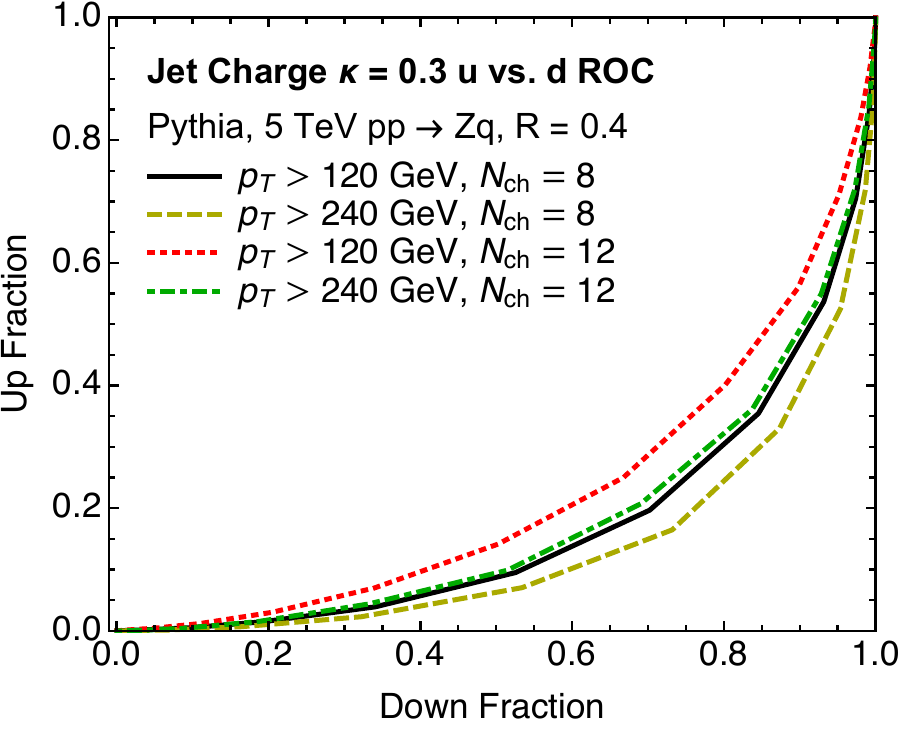}
\caption{\label{fig:roc_pt}
Plot of ROC curves for up versus down quark discrimination from the jet charge with $\kappa =0.3$. Here jet samples are inclusive over charged particle multiplicity (left) or separated into two charged particle multiplicity bins $N_{\rm ch}=8, 12$ (right).}
\end{center}
\end{figure}

Because particle multiplicity increases as $p_T$ increases, our simple scaling also predicts that the discrimination power between up and down jets would decrease as $p_T$ increases.  This is tested at the left of \Fig{fig:roc_pt} where we compare the ROC curves from the jet charge distribution with $\kappa = 0.3$ on jet samples with $p_T > 120$ GeV and $p_T > 240$ GeV.  However, here it is observed that the ROC curve at higher $p_T$ demonstrates better discrimination performance, which is surprising.  Again, the production mechanism of jets is $p_T$ dependent, and apparently an important factor here.  This is probed a bit more on the right plot.  Here, we separate the ROC curve at the different $p_T$s into two bins with $N_\text{ch}=8,12$.  There is a rather significant improvement in discrimination power between samples with the same charged particle multiplicity but different $p_T$, which demonstrates that the jet production mechanism and the way that momentum is shared between charged and neutral particles is important.  However, the difference in ROC curves in the inclusive samples (at left) is much smaller than in the multiplicity bins at right.  This suggests that the increased particle multiplicity at higher $p_T$ is to a large extent reducing the difference in discrimination power.

\begin{figure}[t!]
\includegraphics[width=0.45\textwidth]{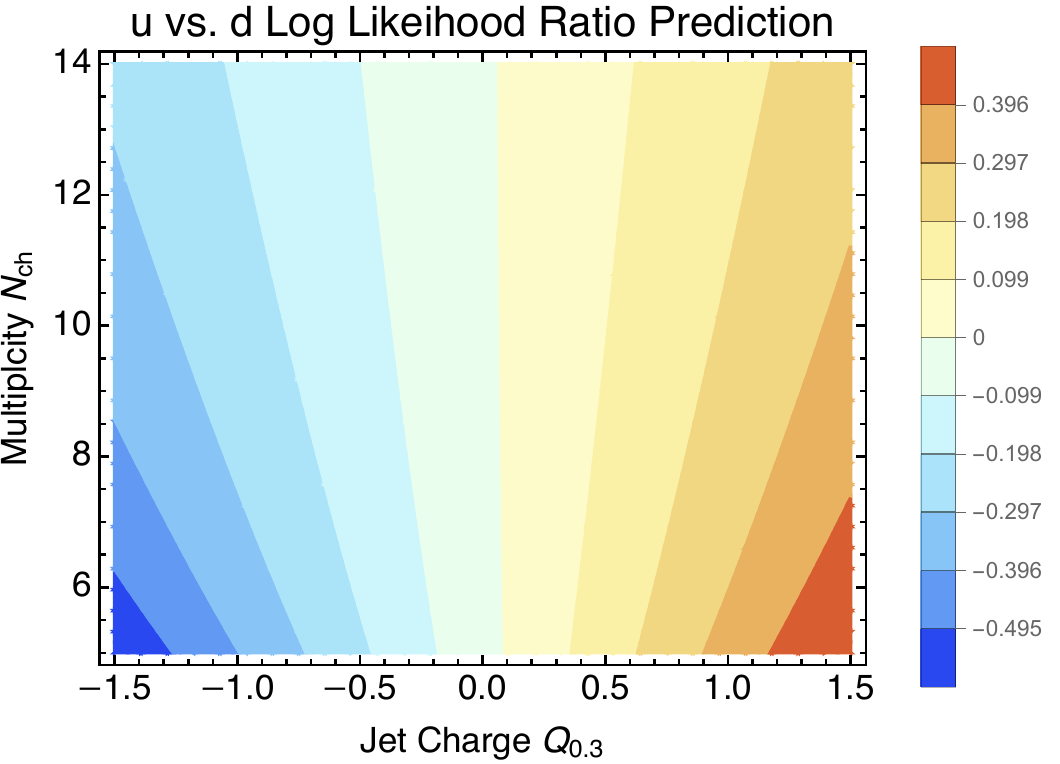}
\caption{\label{fig:loglike_pred}Contour plot of our prediction of the logarithm of the likelihood ratio for up versus down quark discrimination, differential in the jet charge with $\kappa = 0.3$ and the charged particle multiplicity.  Positive (negative) values represent regions dominated by up (down) quark jets.}
\end{figure}

Finally, we plot our scaling prediction for the logarithm of the likelihood ratio, Eq.~(19) in the main text, in \Fig{fig:loglike_pred}.  While the scale of the contours differs, the shape of the contours is in excellent agreement with the output of simulation.  To make this plot, we have set the total particle multiplicity $N = \frac{3}{2}N_\text{ch}$, as expected from isospin.

